# Triage of potential COVID-19 patients from Chest X-ray images using hierarchical convolutional networks

Kapal Dev[1], Sunder Ali Khowaja[2], Ankur Singh Bist[3], Vaibhav Saini[4], and Surbhi Bhatia[5]

[1]CONNECT Centre, Trinity College Dublin; [2]Department of Telecommunication, Faculty of Engineering and Technology, University of Sindh, Jamshoro; Pakistan, [3]Signy Advanced Technologies, India; [4]Indian Institute of Technology, Dehli India; [5]Department of Information Systems in King Faisal University, Saudi Arabia

Email: kdev@tcd.ie & sandar.ali@usindh.edu.pk

Abstract:
The current COVID-19 pandemic has motivated the researchers to use artificial intelligence techniques for a potential alternative to reverse transcription-polymerase chain reaction (RT-PCR) due to the limited scale of testing. The chest X-ray (CXR) is one of the alternatives to achieve fast diagnosis but the unavailability of large-scale annotated data makes the clinical implementation of machine learning-based COVID detection difficult. Another issue is the usage of ImageNet pre-trained networks which does not extract reliable feature representations from medical images. In this paper, we propose the use of hierarchical convolutional network (HCN) architecture to naturally augment the data along with diversified features. The HCN uses the first convolution layer from COVIDNet followed by the convolutional layers from well-known pre-trained networks to extract the features. The use of the convolution layer from COVIDNet ensures the extraction of representations relevant to the CXR modality. We also propose the use of ECOC for encoding multiclass problems to binary classification for improving the recognition performance. Experimental results show that HCN architecture is capable of achieving better results in comparison to the existing studies. The proposed method can accurately triage potential COVID-19 patients through CXR images for sharing the testing load and increasing the testing capacity.

1. Introduction:

CORONAVIRUS is a family of enveloped, positive-sense single-stranded RNA viruses that are important viral pathogens and classified within the Nidovirals order. Considering the history of coronavirus, it has been described for over seven decades, with murine coronavirus JHM being the first to be reported in 1949. In 2003, the variation of the same virus reported 770 deaths from severe acute respiratory syndrome (SARS) [1]. Later in 2004 and 2005, severe bronchiolitis and upper respiratory tract infections were reported to be caused by the family of the same virus. The zoonotic virus MERS, transmitted from animals to humans, another in the list from the same family caused over 850 deaths [2]. Finally, in December 2019, pneumonia-like cases were reported in the Wuhan and later identified as COVID-19. The disease has been declared a pandemic by WHO [3] and is marked as the most severe as any other coronavirus because of its high transmissibility and no preexisting immunity about the virus. The world is undergoing an intimidating situation because of the pandemic novel coronavirus. Currently, over 61.8 million confirmed cases and 1.45 million deaths have been reported as of 28th November 2020, worldwide. The illness comes in the episodes of acute severe respiratory discomfort initiating from throat pain, dry cough followed with a high fever, fatigue, loss of smell or taste, and shortness of breath. The continued progression can lead to the risk of serious medical disorders, such as hypoxia, dyspnea, multiorgan failure, shock, a respiratory failure which requires manual ventilation, and the worst case (death) [4, 5]. As per the center for disease control and prevention (CDC) report, there are many similarities between COVID-19 and flu (influenza), however, they differ in terms of incubation period, symptom onset, shortness of breath, and loss of taste of smell. Furthermore, the severity of the COVID-19 is much higher than the influenza whereas the COVID-19 is more deadly than the later one [6]. The similarity between common flu (influenza) and COVID-19 is another reason to make the testing process faster for its early detection.

The rapid increase in COVID-19 infections is creating a strain on the healthcare facilities around the globe [7, 8]. There are multiple ways for COVID-19 diagnosis but the reverse transcription-polymerase chain reaction (RT-PCR) remains the gold standard that detects the viral nucleic acid from the throat and nasopharyngeal swabs. However, the diagnosis using RT-PCR takes more than 4 – 6 hours and has a low viral load [9] which refers to the amount of virus found in the sputum. Moreover, due to the limited number of test kits, machines, and human experts, the scale of testing is marginally low for developing countries. South Korea is the first one to opt for massive testing but a lot of developed as well as developing countries cannot follow the same strategy due to the above limitations [10–13]. Therefore, a fast way for COVID-19 diagnosis is in need for the timely trace, test, and treat (3T) strategy. Antigen testing is considered to be a fast way of COVID-19 diagnosis but yields poor sensitivity, therefore is not recommended [14].

An alternative to conventional testing is the analysis of chest computed tomography (CT) scans for COVID-19 diagnosis as it belongs to the family of pneumonia. As per the 10th revision of the International Statistical Classification of Diseases and Related Health Problems (ICD-10) COVID-19, SARS, and MERS fall under the category of viral pneumonia whereas the Streptococcus, ARDS, and Pneumocystis belong to the bacterial and fungal pneumonia, respectively [15]. The radiological examination through CT scans has shown better sensitivity in comparison to RT-PCR [9, 16], and can detect COVID-19 from weakly-positive or negative cases declared by RT-PCR. However, despite the better diagnostics, CT-scan has some of the problems similar to that of RT-PCR testing as they are limited in numbers and are too expensive for general masses. Moreover, the CT suites can get contaminated with the regular visits from COVID-19 patients and unlike nasal swabs used in RT-PCR the CT-suites are not disposable, thus, they require extensive cleansing, which put the radiologists and the patients at greater risk [14]. As an alternate modality to CT-scans the chest X-rays (CXRs) have been given a lot of attention for a fast COVID-19 diagnosis. It should be noted that the RT-PCR takes about 4 – 6 hours on average whereas the CXR takes less than half an hour. Furthermore, in case of contaminated input sample, the time gets accumulated which makes the RT-PCR testing highly time consuming.

The ground glass opacities, peripheral, and bilateral consolidation, described by CT-scans can also be reflected by CXR findings [16, 17]. Wong et al. [18] showed that the COVID-19 can be diagnosed using CXRs but yield low sensitivity in comparison to RT-PCR testing. However, there have been some cases (i.e. 9%) where CXRs were able to detect the abnormalities while the RT-PCR tests for those patients were declared negative. It has been established that the CXRs cannot replace the RT-PCR diagnosis at this instance but both of the diagnosing mechanisms can be used to reduce the strain on healthcare systems worldwide. CXR can potentially be used for patient triage as the indication of pneumonia can be detected with higher accuracy. Furthermore, the triage can be extended by distinguishing between bacterial and viral pneumonia so that the RT-PCR resources can be spared, substantially.

The use of bio-inspired artificial intelligence algorithms and deep learning approaches have shown promising results in many fields [19–22] and have been used extensively for the applications such as automatic skin temperature detection, mask detection, social distancing measures, RNA strain analysis, and so forth, during this pandemic. Researchers are actively working on improving the CXR diagnostics for COVID-19 classification. Wang et al. [23] recently proposed COVIDNet from CXR images and have been shown to achieve 91% sensitivity. However, there are a few problems associated with the existing methods. First, the volume of the dataset which is quite limited due to the current public health emergency. Second, the features are either extracted from hand-crafted methods or a single end-to-end deep learning architecture pre-trained on ImageNet which limits the actual representation. Third, due to the high data imbalance, achieving good sensitivity, precision, and accuracy for COVID-19 diagnosis using multiclass classification is quite hard. Fourth, the system is limited in a sense such that it only deals with a few labels, however, the current recognition systems are unable to elevate COVID-19 patients from the patients having flu only. The methods are not capable of determining the severity of the case. Therefore, considering a standalone solution for COVID-19 detection is not possible at this instance.

To overcome the above-limitations we proposed a hierarchical convolutional network (HCN). We solve the data distribution problem by extracting feature maps from multiple pre-trained networks which is a natural way of augmenting images images suggesting that the volume of the data increases in accordance with the number of pre-trained networks used while keeping the number of input images same. We explore both the feature-level and decision-level fusion strategies for the augmentation tasks. The feature representation problem is addressed by using the first convolutional layer of COVIDNet-CXR3C [23] cascading with the initial layers of some of the well-known pre-trained networks to extract the features. We propose the use of ECOC conjunct with HCN to transforms the multiclass into a binary classification problem for improving the classification accuracy and sensitivity. The advantage of using HCN is that it works well with a relatively small dataset and the data augmentation compels the network to avoid overfitting. We present a way to triage potential COVID-19 patients through CXR images with other ways of testing to speed- and scale- up the testing process. We further extend our analysis to compare the performance of HCN variants with the existing works. The class activation maps are also used to show the interpretation of the classification results. The contributions of this study are stated as follows:

1. We propose HCN to classify COVID-19 from CXR images.
2. We propose the use of a unique method, i.e. using the first layer of COVID-19 cascaded with the initial layers of pre-trained networks, to augment the data.
3. We propose the ECOC encoding scheme to transform the multiclass into the binary classification problem.
4. We present a potential triage strategy for speeding and scaling up the testing process.

The rest of the paper is structured as follows. Section 2 provides a literature review of existing works. Section 3 describes the proposed methodology. Section 4 presents the experimental results and comparison with the existing works. This section also proposes a strategy to triage COVID-19 patients. Section 5 concludes the paper.

2. Related Work

There are a lot of research involved for practicing computer assisted systems in the domain of healthcare. Those researches include almost all the areas of medicine and obtain efficacious results especially when dealing with the images that are produced in the domain of medicine. For this purpose, a number of deep learning architectures were analyzed before employing certain known neural nets. Many studies have explored CXRimages as a surveillance tool for diagnosing and screening COVID-19 using deep learning techniques. With relevance to the proposed work, we divide the related works section into three divisions. The first subsection will highlight the studies using deep learning techniques specifically on the CXR modality. The second subsection consolidate the works using CXR images for COVID-19 diagnosis, and the third present the works focusing on feature and decision-level fusion strategies for COVID-19 diagnosis using CXR images. CXRImages using Deep Learning: Rajpurkar et al. [24] developed an algorithm for multiclass classification on CT scan (CXRdataset) i.e. ChestX-ray14 dataset. They built a 121 layers CNN architecture to classify the features of the given input x-ray images to one of the 14 different classes. The algorithm was named as CheXNet reporting an accuracy ranging from 0.73 to 0.93 for all 14 classes. The DeepRadiology Team [25] described an approach to pneumonia detection in chest radiographs. Their method used an open-source deep-learning object detection based on CoupleNet (a fully connected CNN)

The local and global features were extracted with the intent to classify pneumonia. The model's accuracy was improved further using ensemble algorithm. Jakhar et al. [26] focused on diagnosing the presence of Pneumothorax using the frontal view of CXRimages. The segmentation techniques have been used to extract the features and predict an output mask correspondingly. U-Net and Pretrained ResNet weights were used to achieve detection at a very fast and accurate way with promising results. Ranjan el at. [27] used interpolation techniques by down-sampling the high dimensional medical images and further feeding them into the deep learning architecture. An autoencoder is created which includes an encoder, decoder and a CNN classifier to reconstruct the input images. A combination of MSE and BCE loss were used at last to predict the thoracic disease in the compressed domain obtained after autoencoders. Wang et al. [28] studied ChestX-ray8 dataset with 8 different classes of disease as atelectasis, cardiomegaly, effusion, infiltration, mass, nodule, pneumonia, and pneumothorax using deep CNN based on unified weakly-supervised multi-label image classification and disease localization formulation. This dataset was trained and tested on different pre-trained networks like AlexNet, VGG16, GoogleNet, and ResNet-50. COVID-19 Diagnosis with CXR images: Basu and Mitra [29] demonstrated a novel technique using transfer learning for diagnosing COVID cases with chest X-rays. Their algorithm was named as Domain Extension Transfer Learning (DETL) that used the Gradient Class Activation Map for finding the characteristic features from the large dataset from different sources of radiology. The model with visual pattern was efficient for distinguishing between classes of COVID. Ozturk et al. [30] discussed the method using DarkNet model and used YOLO object detection system. The codes were created for assisting the radiologist for initial testing and screening for COVID cases. A deep model using CXR images was proposed with accuracy reported for two findings as (binary) 98.08% and for multiclass cases as 87.02%. Farooq and Hafeez [31] studied the differentiation of the COVID-19 cases from that of the pneumonia cases using the CXRimages. A pre-trained ResNet-50 architecture was evaluated on the two publicly available datasets by using a three step technique. Their work includes the preprocessing steps on images as progressive resizing, cyclical learning rate findings and descriptive learning rate findings. Kumar et al. [32] presented the use of ResNet152 and machine learning classifiers distinguishing between cases of COVID-19 and non COVID-19 on CXRimages. Different classifiers were used for evaluating the performance with an accuracy of 0.973 with Random Forest and 0.977 with XGBoost is reported in the paper. Zhang et al. [33] developed a deep-learning model which was composed of a backbone network, a classification head, and an anomaly detection head. High-level features were generated from images using the backbone network and further these features were passed onto the heads to extract the classification score and anomaly score. Binary cross-entropy loss was used for classification score and deviation loss for anomaly score. Abbas et al. [34] proposed a DeTraC CNN architecture for the classification of the CXRimages as COVID-19. DeTraC is an acronym for Decompose, Transfer and Compose which deals with any irregularities with the help of class boundary obtained by class decomposition method. The extra decomposition layer for adding more flexibility to their decision boundaries was added with the motive to decompose each class into subclasses and assigning labels to the new set of class for getting final results. The shallow tuning mode for feature extraction was employed. The model gave high accuracy of 95.12% with comprehensive dataset of images. Misra et al. [35] presented a multi-channel transfer learning model based on ResNet architecture on different sets of dataset composition. 3 classes as normal, pneumonia and COVID were used as the target values. So, they used 3 subnetwork model of binary classification for them similar as one vs all classification. Further, a fine-tuning with another model to do to achieve the classification output. Wang et al. [23] investigated COVIDNet to makes predictions using an explain ability method based on Deep Neural Networkbased architecture for detection of the COVID disease. The customized lightweight design pattern was an added advantage with reduced computational complexity. Adam optimizer was used using a learning rate policy on different set of datasets, namely publicly available dataset and COVIDx dataset. This dataset was generated which contains 358 COVID-19 affected X-ray images, 8066 normal x-rays and 5538 pneumonia affected x-rays. Both quantitative and qualitative analysis was conducted on the above dataset with good score of about 93.3% test accuracy.

Rodolfo et al. [36] discussed features and decision level fusion method by proposing resampling algorithms for detecting COVID on CXR images. Texture descriptions were used for extracting multiple features with CNN and several fusion techniques were used for strengthening the texture descriptions for final classification. The results of the research tested in RYDLS-20 gave an average results of 0.65 with macro average F-score metric using a multi-class labelled data samples and 0.89, as F-1 score for classifying COVID-19 in the hierarchical classification case. Dhurgam et al. [37] proposed and demonstrated novel approach of extracting features from the LBP-transformed CXR images of different types of Chest infections, with the aim of developing an automatic CAD system to be used for distinguishing between COVID-19 and Non COVID-19. The findings resulted an accuracy of 94.43%. Rahimzadeh and Attar [38] proposed the concatenated neural network by capturing the features extracted from Xception and ResNet50V2 and then fusing it to a convolutional layer that is connected to the classifier. The dataset included 180 CXRof COVID patients, 6054 pneumonia affected X-rays and 8851 normal xray images. The fine-tuned ResNet50V2 and Xception network predicts the result with the average accuracy as 99.56%. The uniqueness of the proposed work can be highlighted from both fronts, i.e. learning representations and network design. We propose a way to naturally augment the data by using first convolutional layer from COVID-19 followed by the well-known pre-trained networks. We show that the resultant representations are more diverse than the ones used in existing studies. Moreover, the proposed way of extracting features uses the ImageNet pre-trained networks effectively in compliance with the transfer learning dynamics. We also propose the use of ECOC for label encoding in order to transform the multiclass into binary classification problem which to the best of our knowledge has not been explored yet. Furthermore, this study explores various fusion strategies using the proposed learning representations and the network architecture so that the strategy attaining best results can be selected for comparative analysis.

3. Proposed Method

The proposed architecture for HCN is shown in Fig 1. The hierarchical networks are opted due to their topological flexibility for handling iterative algorithms and cascaded networks. As mentioned earlier, we use the first convolutional layer of pre-trained COVIDNet followed by the initial layers of well-known pre-trained networks to extract the feature maps. The COVIDNet is an open source deep neural network design for detecting COVID-19 from CXR images. The pooling layers are designed such that they down-sample and up-sample the feature maps depending on the stage they are employed and selects a single response by performing convolution-sum fusion. The maps are then encoded for their labels using the ECOC technique. The ECOC techniques are categorized as meta-learning method which transforms the feature space in hot encoded values and solves the multi-class classification problem through employing multiple binary classifiers. The feature maps along with their encoded labels are then trained using DarkNet19 network architecture. The DarkNet19 architecture serves as a backbone convolutional neural network for YOLOv2, a famous object detection framework. The details for each of the blocks in Fig 1 are provided in the subsequent sections.

3.1 Datasets

We employ the COVIDx dataset proposed in [23] to evaluate the COVID detection performance using HCN. The dataset comprises of 13,975 CXR images obtained from 13,870 patients. The reason for the discrepancy in the number of patients and CXR images is due to the fact that at times multiple images are obtained from the same patient. The COVIDx dataset is the combination of five publicly available dataset repositories such as COVID-19 radiography database [39], RSNA Pneumonia detection challenge dataset [40], ActualMed COVID-19 CXR dataset initiative [41], COVID-19 CXR dataset initiative [42], and COVID-19 image data collection [43]. All the above-mentioned datasets are open to the public and are continuously updated constantly. One of the motivations of this work was to deal with the low volume of CXR images representing COVID patients. The COVIDx dataset has around 5,5328 and 8,066 CXR images for Normal and Bacterial Pneumonia, respectively, whereas only 358 CXR images are available for COVID-19 patients (Viral Pneumonia). The high data imbalance justifies our motivation for using activation maps from multiple pre-trained networks to increase the volume of COVID positive CXR images. We adopt the dataset generation method from [23]. We used the same number of training, validation, and testing images as of COVID-19 so that a fair comparative analysis could be carried out. We follow all the ethical policies and guidelines suggested by the authors of different datasets, accordingly.

3.2 Convolutional Layer from COVIDNet

We made our basis earlier that using the networks pre-trained on ImageNet might not provide us better feature representations that are supported by the research community. Veronika cheplygina [44] conducted a study to provide a stance on whether the use of ImageNet pre-training is useful in medical imaging studies or not. The conclusion of the said study was: "it depends" suggesting that if the volume of the data is small then it's better to use a pre-trained network rather than initializing the weights randomly, however, if the volume of data is enough then the network should be trained on medical images from scratch. Raghu et al. [45] also conducted a similar study and suggested that although the ImageNet pre-training is not beneficial in terms of accuracy and precision it does provide faster convergence. In compliance with the existing studies, we use the first convolutional layer from COVIDNet [23] to process the feature maps. The first layer has a convolutional filter size of 7x7 with a stride of 1x1. The feature dimensions are set to be 48 but the subsequent pooling layer will select a single response (as discussed in the next subsection).

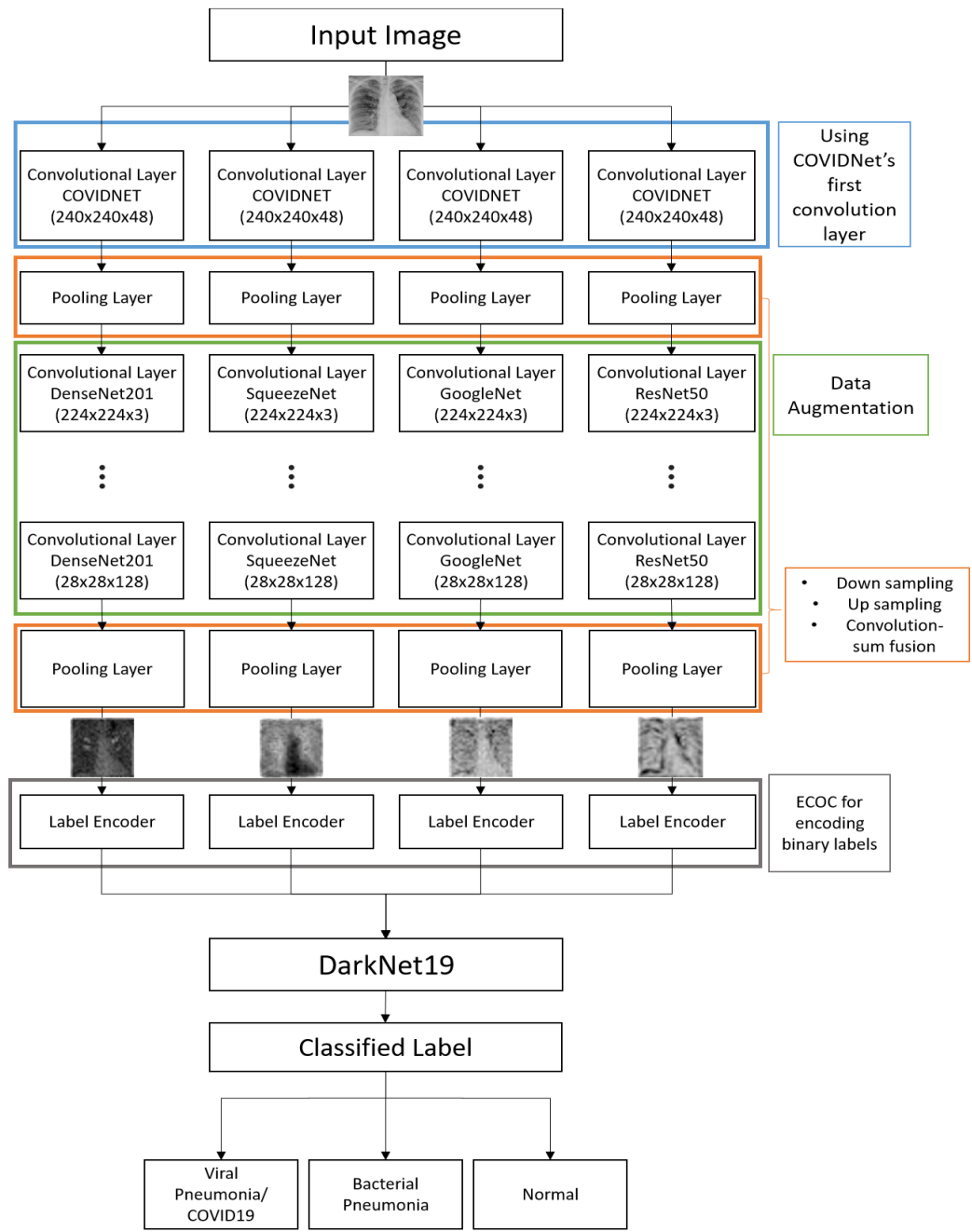

Figure 1 The proposed HCN-FM architecture

As the COVIDNet model was designed by GenSynth, the authors do not provide the architecture code directly. Therefore, we used the checkpoint utilities from tensorflow python training package, along with the index and the pre-trained model (.data) file provided by the authors of COVIDNet [23] to obtain the weights of the first convolutional layer (conv1_conv/kernel). Intuitively, we assume that using the first convolutional layer from COVIDNet provides a better justification of transfer learning than by using standalone network architectures pre-trained on ImageNet (natural and colorful images). As the COVIDNet leverages the principles of residual network architecture, we compare the feature maps extracted from the first convolutional layer of well-known networks designed with residual connections such as ResNet50,

DenseNet201, GoogleNet, and SqueezeNet pre-trained on ImageNet and the one extracted from the first convolutional layer of COVIDNet as shown in Fig 2. To get a single response we use the max pooling layer for the aforementioned convolutional layers, accordingly. The visual difference in the feature maps is quite apparent. The feature map from the pre-trained networks extract some low-level information based on intensity levels which is beneficial for natural and colorful images but does not contribute much to the medical images, whereas the feature map from the COVIDNet's convolution layer focuses on the lung areas which is the cornerstone to detect COVID-19.

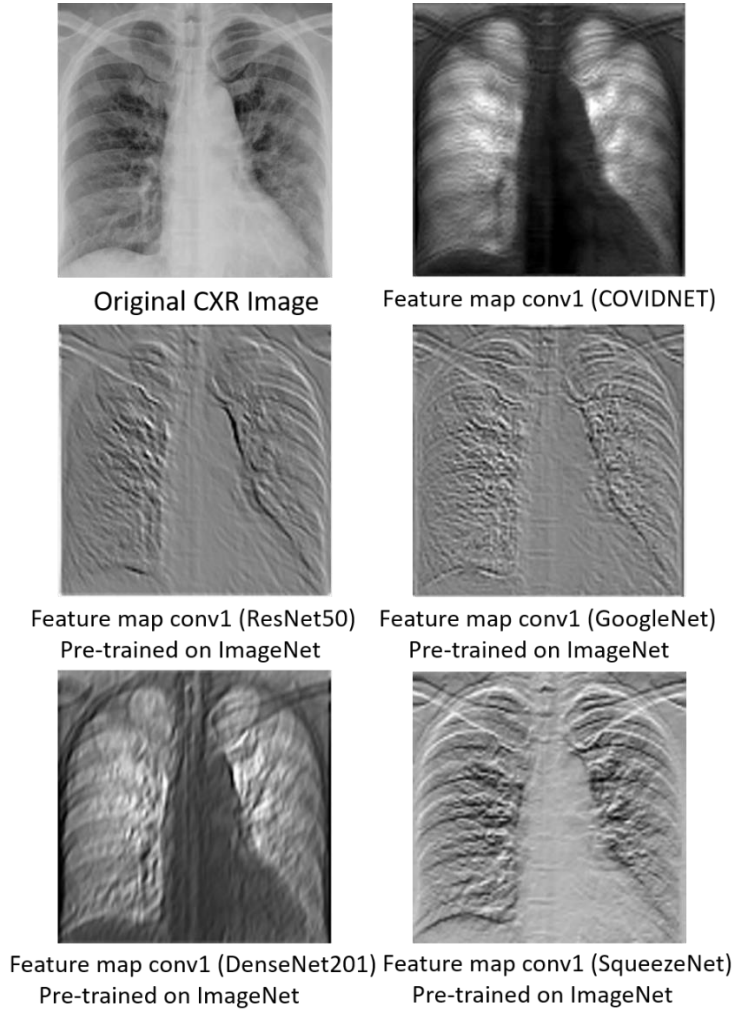

Figure 2 Feature maps extracted from the first convolutional layer of COVIDNet and well known pre-trained networks such as ResNet50, GoogleNet, DenseNet201, SqueezeNet, pre-trained on ImageNet.

3.3 Pooling Layer

In the proposed HCN-FM architecture the first pooling layer is responsible for two operations. The first is the max pooling from the convolutional layer of COVIDNet and the second is the reversible downsampling of the feature maps into three sub-images. Subsequently, the tensor $\ell$ of size $W \times \frac{H}{3} \times C$ is formed and provided as an input to the initial layers of well-known convolutional networks pre-trained on ImageNet where $\ell$ refers to the feature map, and $W$, $H$, and $C$ represent the width, height, and channel, respectively. Zero-padding is performed to keep the size of the feature map compliant with the input size of pre-trained convolutional networks. Similarly, the second pooling layer also comprises of two operations. The first is the fusion of feature maps to provide a single feature response. Studies working on image analysis have extensively applied multiple fusion strategies to improve the recognition performance [46, 47] which include sum, convolution, convolution-sum, and so forth. As suggested in [46, 47] the convolution-sum strategy performs better than both the sum and convolution fusion strategies, therefore, in this study we use the convolution-sum strategy in our second pooling layer. Let $\ell_d \in \mathbb{R}^{H \times W \times D}$ represent the $d^{th}$ feature map where $d = 1, \ldots, D$. The convolution-sum fusion function $f$ performs four steps: (1) concatenation, (2) convolution, (3) dimension reduction, and (4) summation. The feature maps $\ell_d$ and $\ell_{d-1}$ will be concatenated at spatial locations. The convolutional operation is performed through the filter banks which are defined in equation (1).

$$\ell^{conv-sum} = f^{sum}(f^{conv}(\ell_d, \ell_{d-1}), \ell_d)$$

where $f^{sum} = \ell_d + \ell_{d-1}$ and $f^{conv} = (\ell_d \# \ell_{d-1}) * filt + bias$ \qquad (1)

The summation function is just the summation of feature maps with respect to their spatial locations as is denoted by "+". The concatenation operation is represented by "#" which concatenates the features maps horizontally. The convolution operation represented as "*" is applied through the filter banks and biases such that $filt \in \mathbb{R}^{1 \times 1 \times 2w \times h}$ and $bias \in \mathbb{R}^h$, respectively. A weighted combination of feature maps is generated through the convolution operation at spatial location $w \times h$ followed by the reduction in dimension therefore the resultant feature representation will retain the actual size of the map. The second operation executed at this pooling layer is upscaling of feature map performed by the reverse operator from the first pooling layer to produce the map with the same *W*, *H*, and *C*, as the input.

3.4 Initial Convolutional Layers from pre-trained networks

We use initial convolution layers from multiple network architectures pre-trained on ImageNet. Naturally, the question arises as to why we do not use COVIDNet layers as feature extraction? This work aims to extract diverse features to augment the data and in turn increasing the data volume. The problem with using COVIDNet layers as feature extraction is that we lose the diversification of feature maps. Furthermore, the characteristics of COVIDNet in the feature maps are retained from each pre-trained network architecture as the forward pass of the CXR image is propagated through the first convolution layer of the said network. The size of the extracted feature map from each pre-trained network is kept constant by branching off the networks at a specific layer. For instance, ResNet50, DenseNet201, GoogleNet, and SqueezeNet [48–51] are branched off at 38[th], 54[th], 28[th], and 29[th] layer, respectively, to get a 28x28x128 feature responses with added zero-padding to maintain the consistency in image size. The diversified feature maps from the first convolutional layer of the aforementioned pre-trained networks preceding the first convolution layer of COVIDNet is shown in Fig 3.

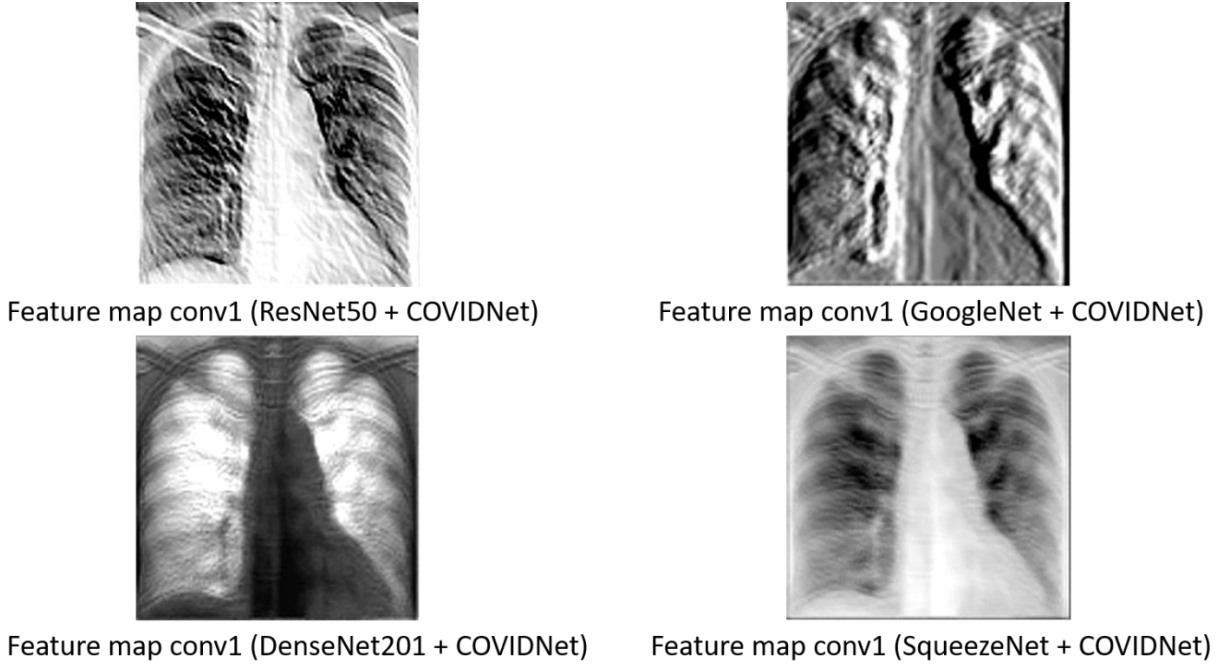

Figure 3 Feature maps from the first convolutional network of the pre-trained networks preceded by the COVIDNet convolutional layer

3.5 Label Encoder

The extracted feature maps will be assigned an encoded label using error-correcting output code (ECOC) in the form of {-1, 0, +1} where 0 represents the non-participating class, -1 and +1 refers to the negative and positive class, respectively. An example of encoded labels for CXR images is shown in Fig 4. The encoded labels will be assigned to each image for each hierarchy. For instance, a feature map will first be assigned either -1 or +1 for the first hierarchy and the darknet19 architecture will be trained on the encoded labels, accordingly. Similarly, the same feature map will undergo the second hierarchy and assigned a label amongst 0, -1, or +1 followed by the darknet19 training. This is how the multi-class problem is transformed into a binary class. Let $c$ define the number of classes and $k$ be the length of the coding matrix, then the encoding matrix $z$ for the ECOC will be of the size $\{-1, 0, +1\}^{c \times k}$. The underlying assumption for using the coding matrix is that each column of $z$ will have $k$ classifiers (binary) which implies that the feature maps $z_{ck}$ comprising of +1 or -1, i.e. positive or negative label, trains $k$ classifiers based on $c - th$ label and the $z_{ck}$ having 0 encoded value will not participate in the training of $k - th$ classifier. We are only interested in the encoded labels assigned to the feature maps by the $k - th$ classifier instead of directly classifying them, the idea is to optimize the separation of feature maps based on their class labels through encodings to represent a binary classification problem. The hierarchical assignment of encoded values will be continued until a leaf node occurs. The weights and biases are characterized as $\mathcal{W}$ and $\mathcal{B}$, respectively. We adopt the joint binary classifier learning (JCL) framework and the optimization method from the studies [52, 53]. The principle optimization problem is shown in equation (2)

$$\min_{\mathcal{W},\mathcal{B},\{\vartheta\},\{\alpha\}} \sum_{n=1}^{N} \delta \sum_{n=1}^{N_k} (\vartheta_n^i) \alpha_n^i + \frac{\lambda}{2} \sum_{i=1}^{k} \xi |\vartheta_i|$$

Subject to:

$$\vartheta_c \in \{-1, 0, +1\}, \forall\, c \in classes,$$

$$\vartheta_n(\mathcal{W}'f_n + \mathcal{B}) \geq 1 - \alpha_n, \forall_n$$

$$\alpha_n \geq 0, \forall_n$$

$$-\tau \leq \sum_1^c \vartheta \leq \tau,$$

$$\sum_{i=1}^{k} 1\{\vartheta_i < 0\} \geq 1 \text{ and } \sum_{i=1}^{k} 1\{\vartheta_i > 0\} \geq 1 \tag{2}$$

The variable $\delta$ refers to the mismatch loss between the actual encoded label and the predicted encoding. The notation $n$ refers to the number of training samples, specifically the number of feature maps used for training. The first constraint in equation 2 refers to the coding variable $\vartheta$ which can take the values $\{-1, 0, +1\}$ representing the label $c$. The variables $\mathcal{W}, \mathcal{B}, \{\alpha\}$ are the optimizable parameters that define the decision boundary as presented in the second constraint. The third constraint represents the hinge loss. The term $\tau$ in the fourth constraint refers to the tolerance level which is introduced to maintain the balance. The last constraint ensures that there is at least one positive and one negative class available. Equation 2 is optimized for the label encodings such that the labels are separable through a decision hyperplane. In the training phase, the label encodings are provided based on the categorization in Fig 3, but the equation 2 is optimized for the testing phase so that the images are assigned the label encodings so that the specific trained model could be activated for classification. For more details regarding the optimization problem, refer to the study [53].

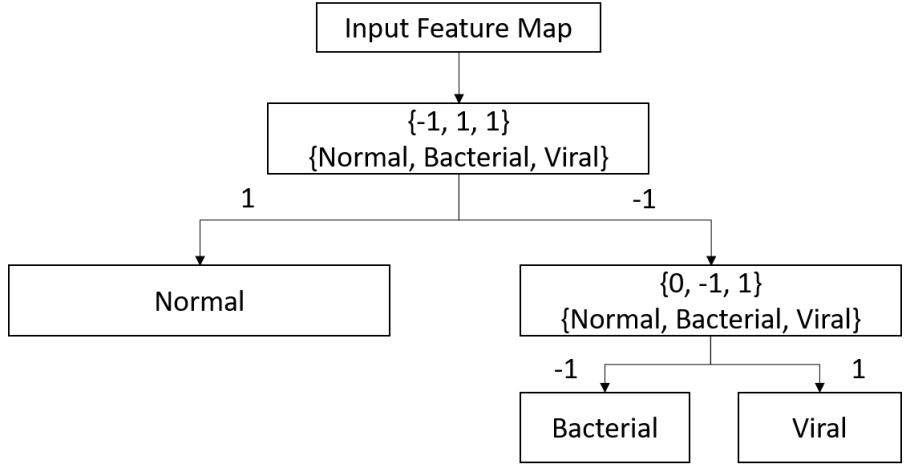

Figure 4 The ECOC method for label encodings for converting multi-class to binary classification problem where +1 and -1 represent the positive and negative class, and the 0 refers to the confusing class ought to be ignored in the training and at the inferential stage.

3.6 DarkNet19 architecture

The reason for choosing DarkNet19 architecture is that it is as accurate as ResNet architectures but takes 4x less time to train [54, 55]. It is apparent from the recent studies on COVID-19 diagnosis that the volume of viral pneumonia CXR images are less in number. Therefore, to balance the distribution of data we use the bootstrapping method with sample replacement at the batch level. We fine-tune the DarkNet19 network pre-trained on ImageNet with ADAM optimizer [56] having an initial learning rate of 0.003, batch size of

32, and an exponential decay rate of 0.00005 after every 3 epochs. We train the network for 40 epochs. We used the rotation, translation, flipping, and zoom augmenting strategies for fine-tuning the network.

3.7 Fusion Strategies

The image analysis studies use fusion strategies extensively to improve recognition performance. We evaluate four different kinds of strategies that vary based on computational complexity and the number of parameters. We evaluate decision-level fusion strategy with meta-learning, decision-level fusion strategy with weighted averaging, feature-level fusion, and without fusion. The strategies are listed in their descending order suggesting that the decision-level fusion strategy with meta-learning comprises a large number of parameters and high computational complexity whereas the training without fusion has the lowest number of parameters and lower computational complexity amongst all the strategies. The decision-level fusion needs separate streams to be trained for encoded labels in a hierarchical fashion suggesting that a separate DarkNet19 network will be trained on features extracted from ResNet50 and so forth. The probabilities from each of the streams will be combined either using a meta-learning strategy [53] or weighted averaging [47]. The feature-level fusion combines the feature maps using gradient-sum pooling [47] and trains a single DarkNet19 architecture for the encoded labels in the proposed hierarchy. We refer to the decision-level fusion with meta-learning as HCN-DML, decision-level fusion with weighted averaging as HCN-DWA, feature-level fusion as HCN-FLF, and without fusion as HCN-FM. The use of the aforementioned fusion strategies with reference to the proposed architecture is shown in Fig 5. The HCN-FM represents a cross-modal training architecture [47] where the diverse features are trained using a single-stream. On the other hand, HCN-DML, and HCN-DWA takes an opposite approach by training an individual stream for a single modality. We adopt the implementation of HCN-DML from [53], HCN-DWA, and HCN-FLF from [47], respectively.

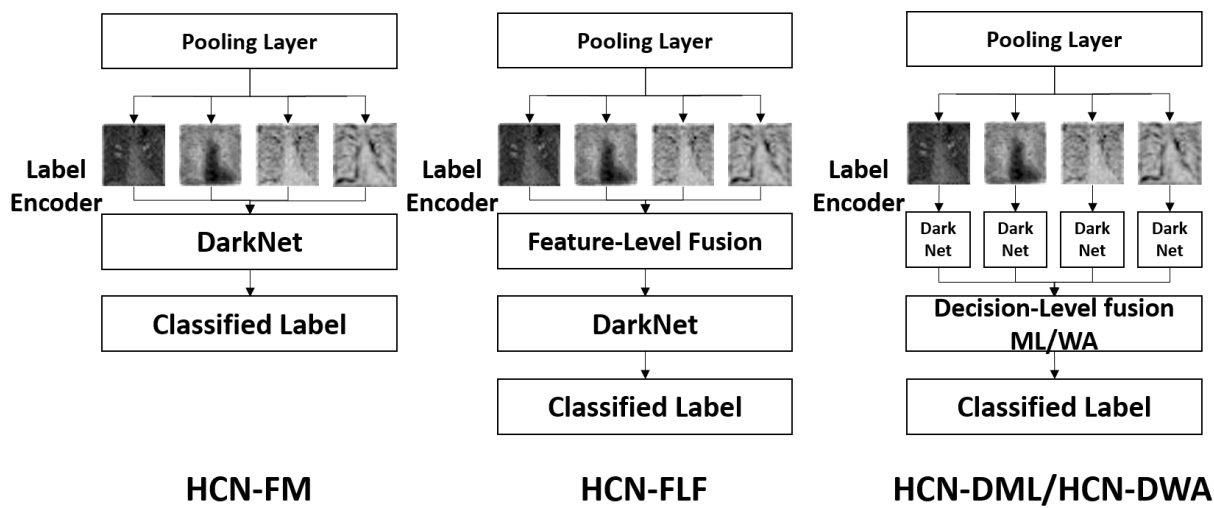

Figure 5 Application of fusion strategies with reference to the proposed architecture. HCN-FM → HCN without fusion, HCN-FLF → HCN with feature-level fusion, the feature maps are fused using gradient-sum pooling [47], HCN-DML → HCN with meta-learning based decision-level fusion, the probabilities of the classified labels are trained using a shallow-learning classifier [53], and HCN-DWA → HCN with weighted averaging based decision-level fusion, the probabilities of the classified labels are combined through weighted averaging and the class with maximum probability is selected.

4. Experimental Results

This section first provides the details for the dataset employed in this study followed by the quantitative and qualitative analysis for evaluating the effectiveness of the proposed HCN-FM. We also perform ablation studies to select the best fusion strategy, accordingly. Furthermore, this section presents an implication of the proposed work in the form of a strategy for the triage of COVID-19 patients to support other testing approaches.

4.1 Ablation Study

As the proposed hierarchical classification network employs various fusion strategies, it is essential to evaluate each of the strategies quantitatively. This will allow us to investigate not only the test accuracy but also the trade-off between the accuracy and computational complexity. We evaluate HCN-DML, HCN-DWA, HCN-FLF, and HCN-FM using sensitivity, specificity, precision, and accuracy metric on test set images from the COVIDx dataset. The results are shown in Table 1. The results show that the HCN-DML achieves the best result in terms of sensitivity, specificity, precision, and accuracy, accordingly. It makes sense as the HCN-DML has the highest computational complexity in comparison to the other strategies. An interesting finding here is that the HCN-FM performs better than the HCN-DWA that performs a weighted averaging on the probabilities obtained using four different network streams. Additionally, the HCN-DWA has the second-highest computational complexity whereas the HCN-FM has the lowest one and the least number of parameters.

Table 1 Quantitative results for the fusion strategies using hierarchical classification network

| Fusion Strategy | Normal | Bacterial Pneumonia | Viral Pneumonia |
|---|---|---|---|
| Sensitivity | | | |
| HCN-DML | 96.08% | 95.10% | 98.96% |
| HCN-DWA | 93.20% | 94.00% | 94.85% |
| HCN-FLF | 91.26% | 90.38% | 96.77% |
| HCN-FM | 91.51% | 95.05% | 100.00% |
| Precision | | | |
| HCN-DML | 98.00% | 97.00% | 95.00% |
| HCN-DWA | 96.00% | 94.00% | 92.00% |
| HCN-FLF | 94.00% | 94.00% | 90.00% |
| HCN-FM | 97.00% | 96.00% | 93.00% |
| Specificity | | | |
| HCN-DML | 96.97% | 97.47% | 95.59% |
| HCN-DWA | 94.42% | 94.00% | 93.60% |
| HCN-FLF | 93.40% | 93.88% | 90.82% |
| HCN-FM | 97.42% | 95.48% | 93.24% |
| Accuracy | | | |
| HCN-DML | 96.67% | | |
| HCN-DWA | 94.00% | | |
| HCN-FLF | 92.67% | | |
| HCN-FM | 95.33% | | |

We assume that the resultant classification probabilities are quite close to each other for individual streams which makes it difficult for a weighted averaging scheme to differentiate them. On the other hand, the HCN-FM represents a cross-modal learning strategy but in this case, the cross-modality is represented at the input level, i.e. the feature maps. It has been shown in the literature that cross-modal learning in some cases benefits from the diverse representations and might achieve better results than the multi-stream networks [47]. The lowest recognition rate is obtained using HCN-FLF which is apparent as it uses a single response for training the network stream. The HCN-FLF does not leverage the natural way of data augmentation as proposed in this study. However, the results obtained using HCN-FLF are on par with the COVIDNet, respectively. It should also be noted that the HCN networks also use a bootstrapping technique to balance the distribution of the samples as well as the ECOC technique to transform the multiclass to the binary class classification problem.

Table 2 Ablation study for highlighting the importance of ECOC and bootstrapping in HCN framework

| Fusion Strategy | Normal | Bacterial Pneumonia | Viral Pneumonia |
|---|---|---|---|
| Sensitivity | | | |
| HCN-DML w/o ECOC | 95.05% | 93.20% | 96.88% |
| HCN-DML w/o Bootstrapping | 95.10% | 96.04% | 97.94% |
| HCN-DML w/o ECOC + Bootstrapping | 94.06% | 92.23% | 95.83% |
| HCN-DML | 96.08% | 95.10% | 98.96% |
| HCN-FM w/o ECOC | 92.08% | 87.74% | 97.85% |
| HCN-FM w/o Boot strapping | 94.06% | 90.38% | 97.89% |
| HCN-FM w/o ECOC + Bootstrapping | 97.00% | 92.00% | 83.00% |
| HCN-FM | 91.51% | 95.05% | 100.00% |
| Precision | | | |
| HCN-DML w/o ECOC | 96.00% | 96.00% | 93.00% |
| HCN-DML w/o Bootstrapping | 97.00% | 97.00% | 95.00% |
| HCN-DML w/o ECOC + Bootstrapping | 95.00% | 95.00% | 92.00% |
| HCN-DML | 98.00% | 97.00% | 95.00% |
| HCN-FM w/o ECOC | 93.00% | 93.00% | 91.00% |
| HCN-FM w/o Boot strapping | 95.00% | 94.00% | 93.00% |
| HCN-FM w/o ECOC + Bootstrapping | 88.20% | 86.80% | 98.80% |
| HCN-FM | 97.00% | 96.00% | 93.00% |

We evaluate the two best networks, i.e. HCN-DML and HCN-FM with and without bootstrapping and ECOC technique to get a deeper understanding. The results are reported in table 2. The results show that the transformation to binary class problem plays a vital role in improving the COVID recognition performance from CXR images. The bootstrapping does help in enhancing the sensitivity and precision but the improvement is not significant. The important aspect of this ablation study is that it shows even without using ECOC and bootstrapping the HCN-DML achieves considerably better sensitivity which highlights the intrinsic properties of the proposed HCN and the usefulness of deriving diversified representation. However, the HCN network has been designed to tackle the challenge of detecting COVID-19 from CXR images when using fine-grained labels, such as Streptococcus, Legionella, Pneumocystis, Klebsiella under the category of bacterial pneumonia, and SARS, MERS, ARDS, COVID-19 within the viral pneumonia category. The ECOC has proved to be beneficial in improving the performance with a large number of labels [53], although, the increase in performance is not significant using 3 labels with respect to the HCN-DML but still the encodings improve the performance rather than degrading it. It is also to be noted that the use of ECOC and bootstrapping is supported by the results from HCN-FM where the absence of these techniques results in reduced sensitivity, i.e. 83.00% for viral pneumonia. The results highlight that the ECOC and bootstrapping technique compliment each other for improving the recognition performance. It should also be noted that the sensitivity for almost all the HCN variants is above 90% which is quite acceptable considering that the clinical experts achieved 69% for the same when diagnosing COVID-19 from CXR images. Moreover, the sensitivity of RT-PCR which is currently the gold standard has been recorded to be 91% [18].

4.2 Qualitative Analysis

We illustrate the visualization of saliency maps using the Grad-CAM method [57] in Fig 6 and 7, respectively. The figures 6 and 7 shows the broad main lesion learned by the network to classify COVID-19 patients, correctly. It was noticed that the particular map is only activated in COVID-19 CXR images whereas no saliency map was observed with Bacterial or Normal CXR images as shown in Fig 8. The images support our previous analysis where HCN achieves better sensitivity analysis than the gold standard. Moreover, the Grad-CAM maps can also be used for the interpretability of the CXR images while providing insights to the radiologists through main lesions which might be helpful for clinical diagnosis.

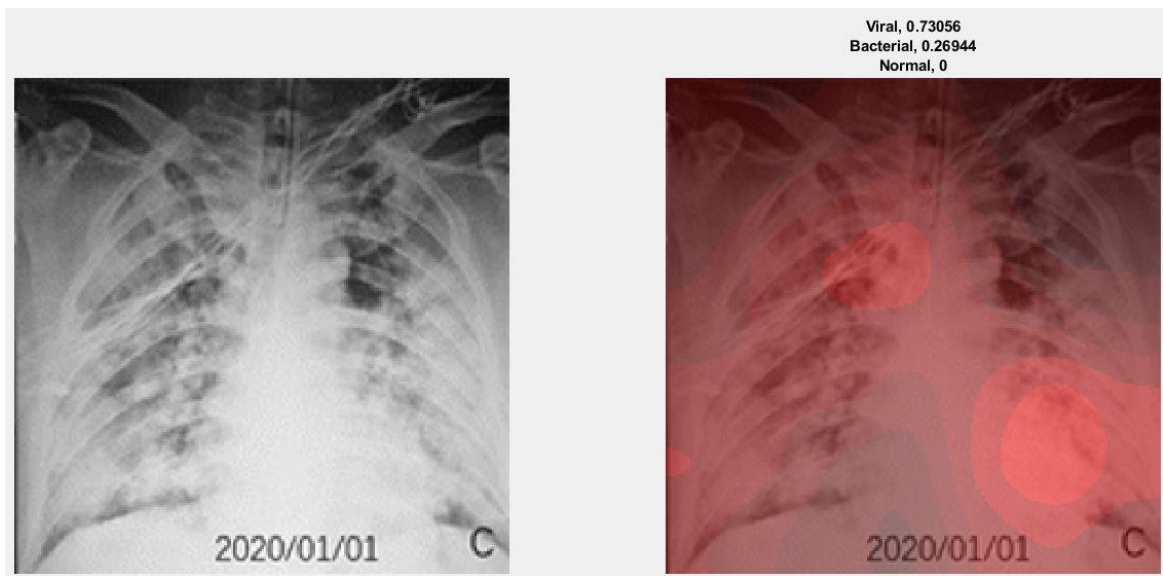

Figure 6 Example of an activation map for Viral Pneumonia patient using Grad-CAM. The left image represents the original CXR whereas the right image represents the saliency map along with classification probabilities

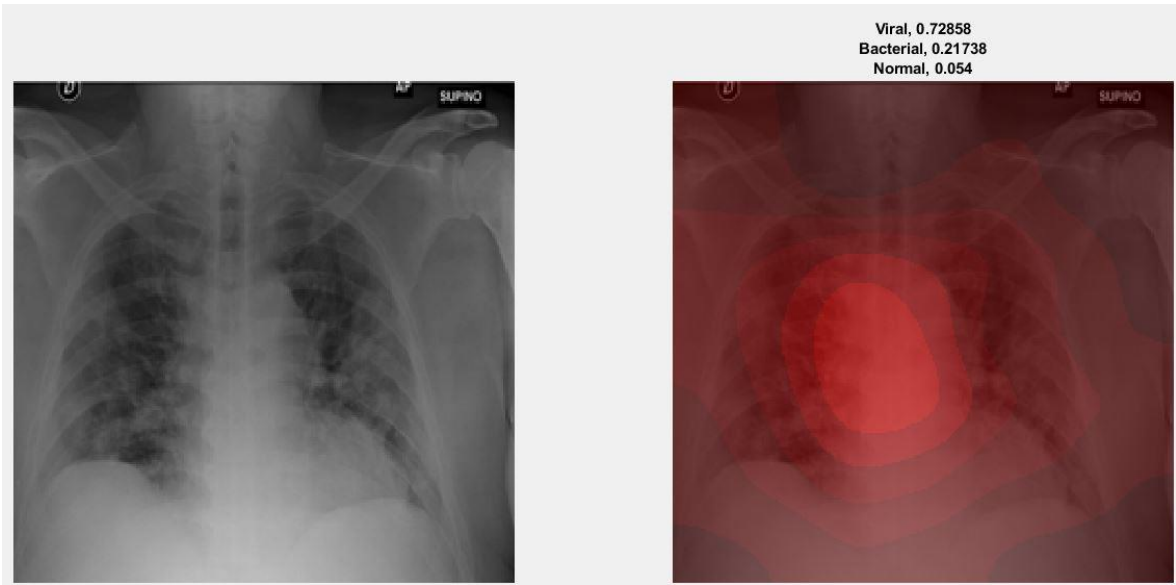

Figure 7 Example of an activation map for Viral Pneumonia patient using Grad-CAM. The left image represents the original CXR whereas the right image represents the saliency map along with classification probabilities

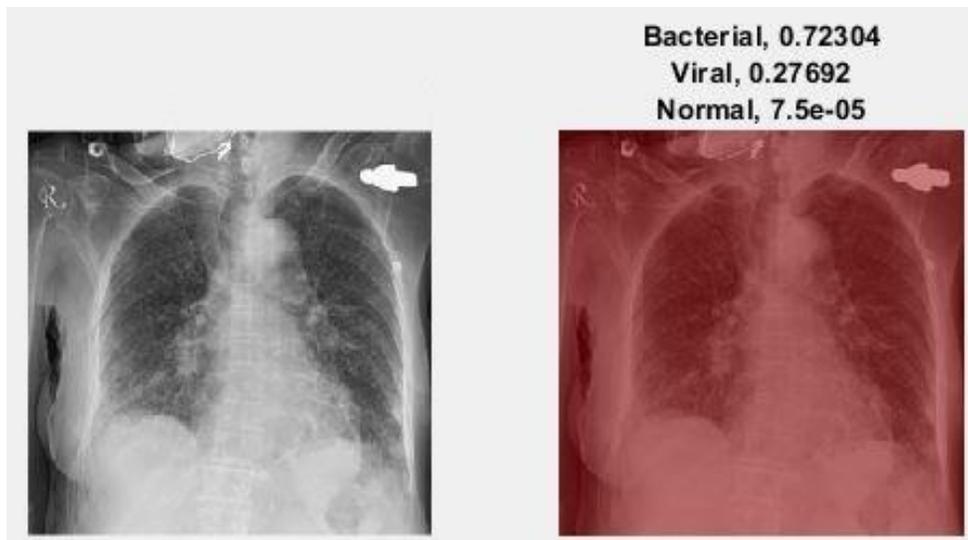

Figure 8 Example of an activation map for Bacterial Pneumonia patient using Grad-CAM. The left image represents the original CXR whereas the right image represents the saliency map along with classification probabilities.

4.3 Comparison with the existing works

All of the experiments reported in this study have been conducted on a GPU NVIDIA GeForce GTX 1080Ti with 32GB RAM and Intel Core i7 clocked at 3.4 GHz. The COVID pandemic is recent therefore, the studies centered towards diagnosing COVID-19 from a limited amount of annotated images are quite less. We compare the proposed HCN architecture with three of the studies that are considered to be state-of-the-art approaches. The first one is the COVIDNet whereas the second and third are proposed in [14] and [58]. The comparison between the HCN architecture and the state-of-the-art approaches is shown in table 3. It should be noted that we used the same dataset along with the same training and testing protocols to perform a fair comparison. The overall accuracy achieved by HCN-FM and HCN-DLM is 96.67% and 95.33% which outperforms the results reported in [14], [58] and COVIDNet, i.e. 91.9%, 94.72%, and 93.33%, respectively. Furthermore, the proposed method shows significant improvement in terms of both, the sensitivity and the precision, accordingly. The training time for HCN-FM and HCN-DLM was noted to be 1055, and 2926 seconds, whereas the testing time for both the networks, was noted to be 3, and 8 seconds, respectively. We do agree that the proposed HCN architecture is more computationally complex than that of [14], [58], and COVIDNet but considering the inference time, reliability of clinical diagnosis, and the ongoing pandemic situation, we believe that the accuracy weighs more than the computational complexity. Some studies proposed the COVID-19 detection from CXR images using the COVIDx dataset but their evaluation is either based on the same protocol as of [23] or they perform the binary classification, i.e. COVID-19 and non-COVID-19 patients. We compare the performance of HCN with the other recent studies in terms of accuracy in table 4. It is evident that the proposed HCN architecture outperforms the recent studies for detecting COVID-19 from CXR images.

Table 3 Comparison of proposed HCN architecture with state-of-the-art methods

| Method | Normal | Bacterial Pneumonia | Viral Pneumonia |
|---|---|---|---|
| Sensitivity | | | |
| COVIDNet [23] | 90.48% | 91.26% | 98.91% |
| Oh et al. [14] | 90.00% | 93.00% | **100.00%** |
| MobileNetv2 [58] | 94.26% | 93.65% | 98.66% |
| HCN-FM | 91.51% | 95.05% | **100.00%** |
| HCN-DLM | **96.08%** | **95.10%** | 98.96% |
| Precision | | | |
| COVIDNet [23] | 95.00% | 94.00% | 91.00% |
| Oh et al. [14] | 95.70% | 90.30% | 76.90% |
| MobileNetv2 [58] | 97.96% | 96.13% | 83.71% |
| HCN-FM | 97.00% | 96.00% | 93.00% |
| HCN-DLM | **98.00%** | **97.00%** | **95.00%** |

Table 4 Comparison with the recent studies in terms of accuracy

| Method | Accuracy |
|---|---|
| COVIDNet [23] | 93.33% |
| EfficientNet [59] | 69.95% |
| ConfiNet [59] | 68.08% |
| AnoDet [59] | 73.24% |
| CAAD [59] | 72.77% |
| DeTraC-ResNet18 [34] | 95.12% |
| VGG19 [60] | 90.00% |
| DenseNet201 [60] | 90.00% |
| MobileNetv2 [58] | 94.72% |
| Hall [61] | 89.20% |
| **HCN-FM** | **95.33%** |
| **HCN-DML** | **96.67%** |

4.4 Triage of HCN for COVID-19

As discussed earlier in this study, the limitation of the gold standard testing, i.e. RT-PCR is due to the shortage of testing kits and expert resources, especially in the developing countries. Keeping in view the current spread of the COVID pandemic, the developed countries also face the shortage of testing kits and medical capacity. It is therefore a need to explore all possible resources and distribute them based on 'triage'. The patients with cough or mild fever rush to the testing facility to get themselves diagnosed. It has been explored by existing studies that people with such symptoms mostly suffer from bacterial pneumonia [62]. The study [62] also suggests that viral pneumonia stands at the third common cause of pneumonia preceded by the S. pneumonia and H. influenza, respectively. For some geological regions, tuberculosis is also considered to be a common cause of pneumonia [62]. The takeaway from the study [62] is that the major portion of patients suffering from the flu, cough, or mild fever might have bacterial pneumonia or tuberculosis rather than COVID-19. Conducting their tests through the gold standard RT-PCR will be a waste of time as well as resources that can be effectively managed through the CXR image diagnosis.

Based on the HCN architecture we present a triage workflow that could help distribute and manage the limited resources during this pandemic in Fig 9. The diagnosis through CXR images is much faster than that of RT-PCR which could be leveraged using HCN architecture. Specifically, the proposed method can distinguish between patients having normal, bacterial, and viral pneumonia. The ones diagnosed with viral pneumonia can be considered for further testing through RT-PCR or CT scans to double-check or validate the diagnosis. This will not only save the medical resources but will also help in speeding up the diagnosis and early isolation of the suspected cases, hence slowing down the spread of COVID-19.

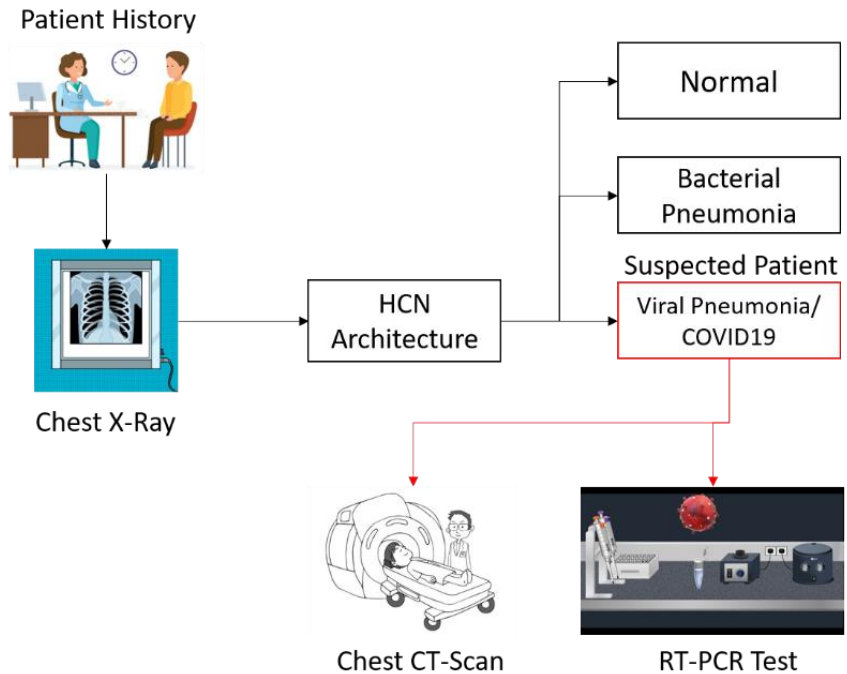

Figure 9 Triage workflow using HCN architecture

5. Discussion and Conclusion

With the current spread of COVID-19 pandemic, the efficient utilization of medical resources is an important issue. The gold standard RT-PCR testing at a large scale is not possible for developing or developed countries. The patient management based on triage is the only option to test the patients and early isolation of the suspects. The CXR and CT scans are potential alternatives to RT-PCR testing. Unfortunately, CT scans suffer from the same problem as of RT-PCR which leaves us with CXR images. To make the diagnosis faster, the use of artificial intelligence can be leveraged by analyzing the CXR images in an automated way. However, the use of artificial intelligence requires a high volume of annotated data which is currently the main problem related to CXR image diagnosis.

In this regard, we proposed the HCN architecture which uses the first convolution layer from COVIDNet followed by the well-known pre-trained architectures to extract the feature representations. This is a natural way to augment the annotated data. Furthermore, we proposed the use of ECOC to transform the multiclass into a binary classification problem for improving the recognition performance. We also performed an in-depth analysis based on the fusion strategies to select the HCN variant with the best recognition performance. The results from the proposed HCN architecture were also verified through qualitative analysis (Grad-CAM) which showed that the activations in the saliency maps are triggered only in the COVID-19 patients' CXR. It should be noted that the proposed work is not designed to elevate COVID-19 from flu, rather the HCN differentiates between viral pneumonia, bacterial pneumonia, and normal CXR images. As per CDC guidelines [6], only 15% of the patients with COVID-19 develop severe symptoms which leads to pneumonia but the same body characterizes COVID-19 in the family of viral pneumonia. However, it can observed from the HCN's Grad-CAM that it only triggers the activation in the saliency map when the COVID (viral pneumonia) is detected rather than the bacterial or fungal infection. The categorization of the bacterial and viral pneumonia in our study is in line with ICD10, thus, we consider the categorization justified. Furthermore, we compared the proposed work with existing state-of-the-art works and showed that the HCN yields better recognition performance. Finally, a triage workflow has also been laid out for showing the real-world applicability of the proposed work.

Even though we report the state-of-the-art results, the proposed work has some limitations for its applicability to the real-world setting. The first limitation is regarding the availability of the limited labeled dataset. Although, this work proposes a natural way of augmenting the data still the dataset is too small to be applied to real-world settings. The second limitation is the label constraints with respect to the dataset. The images are labeled as normal, bacterial, and viral pneumonia, however, there are many classifications within these three labels. For instance, bacterial pneumonia includes Streptococcus, Legionella, Pneumocystis, Klebsiella, and more whereas viral pneumonia includes SARS, MERS, COVID-19, and ARDS. The current system is not designed either to elevate COVID-19 patients from the ones having flu or to classify the severity of the cases. It is in line with the previous limitation, i.e. limited labeled data, accordingly.

One of the benefits of using HCN is the ability to deal with multiple labels by leveraging the concepts of ECOC. As per the medical community, the diseases such as Pleural effusion, Pulmonary edema, Pulmonary fibrosis, and Chronic obstructive pulmonary disease (COPD) yield the same characteristics of lung shrinking in terms of CXR visualization, which can be added to the bacterial variant of pneumonia. However, the point is, there is a need for large-scale annotation with extended labels. As the proposed HCN uses ECOC, it can be helpful to build a recognition system with a large set of labels for differentiating amongst several diseases. As our future work, we want to coordinate with experts to build a database of such diseases to not only detect the disease but also to rate the severity of each disease through CXR images.

References

1. Skowronski DM, Astell C, Brunham RC, et al (2005) Severe Acute Respiratory Syndrome (SARS): A Year in Review. Annu Rev Med 56:357–381. https://doi.org/10.1146/annurev.med.56.091103.134135

2. Song Z, Xu Y, Bao L, et al (2019) From SARS to MERS, Thrusting Coronaviruses into the Spotlight. Viruses 11:59. https://doi.org/10.3390/v11010059

3. (2020) Coronavirus disease (COVID-19) Pandemic. In: World Heal. Organ. https://www.who.int/emergencies/diseases/novel-coronavirus-2019. Accessed 22 May 2020

4. Adhikari SP, Meng S, Wu Y-J, et al (2020) Epidemiology, causes, clinical manifestation and diagnosis, prevention and control of coronavirus disease (COVID-19) during the early outbreak period: a scoping review. Infect Dis Poverty 9:29. https://doi.org/10.1186/s40249-020-00646-x

5. Salehi S, Abedi A, Balakrishnan S, Gholamrezanezhad A (2020) Coronavirus Disease 2019 (COVID-19): A Systematic Review of Imaging Findings in 919 Patients. Am J Roentgenol 1–7. https://doi.org/10.2214/AJR.20.23034

6. (2020) Similarities and Differences between Flu and COVID-19. In: Centers Dis. Control Prev. https://www.cdc.gov/flu/symptoms/flu-vs-covid19.htm. Accessed 29 Nov 2020

7. Bhattacharya S, Reddy Maddikunta PK, Pham Q-V, et al (2020) Deep learning and medical image processing for coronavirus (COVID-19) pandemic: A survey. Sustain Cities Soc 102589. https://doi.org/10.1016/j.scs.2020.102589

8. Singh M, Aujla GS, Bali RS, et al (2020) Blockchain-enabled secure communication for drone delivery. In: Proceedings of the 2nd ACM MobiCom Workshop on Drone Assisted Wireless Communications for 5G and Beyond. ACM, New York, NY, USA, pp 25–30

9. Xie X, Zhong Z, Zhao W, et al (2020) Chest CT for Typical 2019-nCoV Pneumonia: Relationship to Negative RT-PCR Testing. Radiology 200343. https://doi.org/10.1148/radiol.2020200343

10. Engelberg S, Song L, DePillis L (2020) How South Korea Scaled Coronavirus Testing While the U.S. Fell Dangerously Behind. In: ProPublica. https://www.propublica.org/article/how-south-korea-scaled-coronavirus-testing-while-the-us-fell-dangerously-behind. Accessed 30 Jun 2020

11. Bedingfield W (2020) What the world can learn from South Korea's coronavirus strategy. In: WIRED. https://www.wired.co.uk/article/south-korea-coronavirus. Accessed 30 Jun 2020

12. McCurry J (2020) Test, Trace, contain: How South Korea flattened its coronavirus curve. In: Guard. https://www.theguardian.com/world/2020/apr/23/test-trace-contain-how-south-korea-flattened-its-coronavirus-curve. Accessed 30 Jun 2020

13. Sakellaropoulou R (2020) Research in the time of a pandemic: Korea's COVID-19 success story. In: Springer Nat. https://www.springernature.com/gp/researchers/the-source/blog/blogposts-communicating-research/korea-s-covid-19-success-story-/18036412. Accessed 30 Jun 2020

14. Oh Y, Park S, Ye JC (2020) Deep Learning COVID-19 Features on CXR using Limited Training Data Sets. IEEE Trans Med Imaging 1–1. https://doi.org/10.1109/TMI.2020.2993291

15. Association AMe (2018) ICD-10-CM 2019 The Complete Official Codebook

16. Fang Y, Zhang H, Xie J, et al (2020) Sensitivity of Chest CT for COVID-19: Comparison to RT-PCR. Radiology 200432. https://doi.org/10.1148/radiol.2020200432

17. Ai T, Yang Z, Hou H, et al (2020) Correlation of Chest CT and RT-PCR Testing in Coronavirus Disease 2019 (COVID-19) in China: A Report of 1014 Cases. Radiology 200642. https://doi.org/10.1148/radiol.2020200642

18. Wong HYF, Lam HYS, Fong AH-T, et al (2019) Frequency and Distribution of Chest Radiographic Findings in COVID-19 Positive Patients. Radiology 201160. https://doi.org/10.1148/radiol.2020201160

19. Jindal A, Aujla GS, Kumar N, et al (2018) SeDaTiVe: SDN-Enabled Deep Learning Architecture for Network Traffic Control in Vehicular Cyber-Physical Systems. IEEE Netw 32:66–73. https://doi.org/10.1109/MNET.2018.1800101

20. Aujla GS, Jindal A, Chaudhary R, et al (2019) DLRS: Deep Learning-Based Recommender System for Smart Healthcare Ecosystem. In: ICC 2019 - 2019 IEEE International Conference on Communications (ICC). IEEE, pp 1–6

21. Raja G, Manaswini Y, Vivekanandan GD, et al (2020) AI-Powered Blockchain - A Decentralized Secure Multiparty Computation Protocol for IoV. In: IEEE INFOCOM 2020 - IEEE Conference on Computer Communications Workshops (INFOCOM WKSHPS). IEEE, pp 865–870

22. Zehra SS, Qureshi R, Dev K, et al (2020) Comparative Analysis of Bio-Inspired Algorithms for Underwater Wireless Sensor Networks. Wirel Pers Commun. https://doi.org/10.1007/s11277-020-07418-8

23. Wang L, Wong A (2020) COVID-Net: A Tailored Deep Convolutional Neural Network Design for Detection of COVID-19 Cases from Chest X-Ray Images. https://doi.org/2003.09871

24. Rajpurkar P, Irvin J, Zhu K, et al (2017) CheXNet: Radiologist-Level Pneumonia Detection on Chest X-Rays with Deep Learning

25. The DeepRadiology Team (2018) Pneumonia Detection in Chest Radiographs

26. Jakhar K, Bajaj R, Gupta R (2019) Pneumothorax Segmentation: Deep Learning Image Segmentation to predict Pneumothorax

27. Ranjan E, Paul S, Kapoor S, et al (2018) Jointly Learning Convolutional Representations to Compress Radiological Images and Classify Thoracic Diseases in the Compressed Domain. In: Proceedings of the 11th Indian Conference on Computer Vision, Graphics and Image Processing. ACM, New York, NY, USA, pp 1–8

28. Wang X, Peng Y, Lu L, et al (2017) ChestX-Ray8: Hospital-Scale Chest X-Ray Database and Benchmarks on Weakly-Supervised Classification and Localization of Common Thorax Diseases. In: IEEE Conference on Computer Vision and Pattern Recognition (CVPR). IEEE, pp 3462–3471

29. Basu S, Mitra S, Saha N (2020) Deep Learning for Screening COVID-19 using Chest X-Ray Images

30. Ozturk T, Talo M, Yildirim EA, et al (2020) Automated detection of COVID-19 cases using deep neural networks with X-ray images. Comput Biol Med 121:103792. https://doi.org/10.1016/j.compbiomed.2020.103792

31. Farooq M, Hafeez A (2020) COVID-ResNet: A Deep Learning Framework for Screening of

COVID19 from Radiographs

32. Kumar R, Arora R, Bansal V, et al (2020) Accurate Prediction of COVID-19 using Chest X-Ray Images through Deep Feature Learning model with SMOTE and Machine Learning Classifiers. medRXiv. https://doi.org/10.1101/2020.04.13.20063461

33. Zhang J, Xie Y, Pang G, et al (2020) Viral Pneumonia Screening on Chest X-ray Images Using Confidence-Aware Anomaly Detection

34. Abbas A, Abdelsamea MM, Gaber MM (2020) Classification of COVID-19 in chest X-ray images using DeTraC deep convolutional neural network. https://doi.org/10.1101/2020.03.30.20047456

35. Misra S, Jeon S, Lee S, et al (2020) Multi-Channel Transfer Learning of Chest X-ray Images for Screening of COVID-19. Electronics 9:1388. https://doi.org/10.3390/electronics9091388

36. Pereira RM, Bertolini D, Teixeira LO, et al (2020) COVID-19 identification in chest X-ray images on flat and hierarchical classification scenarios. Comput Methods Programs Biomed 194:105532. https://doi.org/10.1016/j.cmpb.2020.105532

37. Al-karawi D, Al-Zaidi S, Polus N, Jassim S (2020) AI based Chest X-Ray (CXR) Scan Texture Analysis Algorithm for Digital Test of COVID-19 Patients. medRXiv

38. Rahimzadeh M, Attar A (2020) A modified deep convolutional neural network for detecting COVID-19 and pneumonia from chest X-ray images based on the concatenation of Xception and ResNet50V2. Informatics Med Unlocked 19:100360. https://doi.org/10.1016/j.imu.2020.100360

39. Rahman T, Chowdhury M, Khandakar A (2020) COVID-19 Radiography Database. In: Kaggle. https://www.kaggle.com/tawsifurrahman/covid19-radiography-database. Accessed 13 May 2020

40. America RS of N (2018) RSNA Pneumonia Detection Challenge. In: Kaggle. https://www.kaggle.com/c/rsna-pneumonia-detection-challenge/data. Accessed 13 May 2020

41. ActualMed (2020) ActualMed COVID-19 Chest X-ray Dataset Initiative. In: GitHub. https://github.com/agchung/Actualmed-COVID-chestxray-dataset. Accessed 13 May 2020

42. Chung A (2020) COVID-19 chest X-ray Data Initiative. In: GitHub

43. Cohen JP, Morrison P, Dao L (2020) COVID-19 Image Data Collection

44. Cheplygina V (2019) Cats or CAT scans: Transfer learning from natural or medical image source data sets? Curr Opin Biomed Eng 9:21–27. https://doi.org/10.1016/j.cobme.2018.12.005

45. Raghu M, Zhang C, Kleinberg J, Bengio S (2019) Transfusion: Understanding Transfer Learning for Medical Imaging. In: Advances in Neural information Processing Systems 32. pp 1–11

46. Feichtenhofer C, Pinz A, Zisserman A (2016) Convolutional Two-Stream Network Fusion for Video Action Recognition. In: 2016 IEEE Conference on Computer Vision and Pattern Recognition (CVPR). IEEE, pp 1933–1941

47. Khowaja SA, Lee S-L (2019) Hybrid and hierarchical fusion networks: a deep cross-modal learning architecture for action recognition. Neural Comput Appl. https://doi.org/10.1007/s00521-019-04578-y

48. He K, Zhang X, Ren S, Sun J (2016) Deep Residual Learning for Image Recognition. In: 2016 IEEE Conference on Computer Vision and Pattern Recognition (CVPR). IEEE, pp 770–778

49. Huang G, Liu Z, Van Der Maaten L, Weinberger KQ (2017) Densely Connected Convolutional

Networks. In: 2017 IEEE Conference on Computer Vision and Pattern Recognition (CVPR). IEEE, pp 2261–2269

50. Szegedy C, Wei Liu, Yangqing Jia, et al (2015) Going deeper with convolutions. In: 2015 IEEE Conference on Computer Vision and Pattern Recognition (CVPR). IEEE, pp 1–9

51. Iandola FN, Han S, Moskewicz MW, et al (2016) SqueezeNet: AlexNet-level accuracy with 50x fewer parameters and <0.5MB model size

52. Tianshi Gao, Koller D (2011) Discriminative learning of relaxed hierarchy for large-scale visual recognition. In: 2011 International Conference on Computer Vision. IEEE, pp 2072–2079

53. Khowaja SA, Yahya BN, Lee S-L (2017) Hierarchical classification method based on selective learning of slacked hierarchy for activity recognition systems. Expert Syst Appl 88:165–177. https://doi.org/10.1016/j.eswa.2017.06.040

54. MC.AI (2018) YOLO3: A Huge Improvement. In: MC.AI. https://mc.ai/yolo3-a-huge-improvement/. Accessed 23 May 2020

55. Joseph R Darknet: Open Source Neural Networks in C. In: https://pjreddie.com/darknet. https://pjreddie.com/darknet. Accessed 23 May 2020

56. Kingma DP, Ba J (2014) Adam: A Method for Stochastic Optimization

57. Selvaraju RR, Cogswell M, Das A, et al (2017) Grad-CAM: Visual Explanations from Deep Networks via Gradient-Based Localization. In: 2017 IEEE International Conference on Computer Vision (ICCV). IEEE, pp 618–626

58. Apostolopoulos ID, Mpesiana TA (2020) Covid-19: automatic detection from X-ray images utilizing transfer learning with convolutional neural networks. Phys Eng Sci Med 43:635–640. https://doi.org/10.1007/s13246-020-00865-4

59. Zhang J, Xie Y, Liao Z, et al (2020) Viral Pneumonia Screening on Chest X-ray Images Using Confidence-Aware Anomaly Detection

60. Hemdan EE-D, Shouman MA, Karar ME (2020) COVIDX-Net: A Framework of Deep Learning Classifiers to Diagnose COVID-19 in X-Ray Images

61. Hall LO, Paul R, Goldgof DB, Goldgof GM (2020) Finding Covid-19 from Chest X-rays using Deep Learning on a Small Dataset

62. Brown PD, Lerner SA (1998) Community-acquired pneumonia. Lancet 352:1295–1302. https://doi.org/10.1016/S0140-6736(98)02239-9